\documentstyle[11pt,aaspp4,tighten]{article}
\begin{document}

\title{\bf Properties of Candidate Cluster Galaxies at $z$=1--2}

\author{Pat Hall}
\affil{Astronomy~Department,~University~of~Toronto,
60~St.~George~St.,~Toronto~ON~M5S~3H8~Canada}

\begin{abstract} 
\noindent{I discuss the properties of a population of candidate cluster galaxies
recently identified in the fields of radio-loud quasars at $z$=1--2.
The magnitude, spatial, and color distributions of the excess galaxies
in these fields are consistent with the quasars residing in moderately
rich galaxy overdensities dominated by early-type galaxies.
The broadband SEDs of very red galaxies in several fields 
suggest that there is considerable dispersion in the properties of
early-type galaxies in these clusters at $z$$\sim$1.5.
Most excess galaxies are best fit as 2--3~Gyr old ellipticals, but some are either
4--5~Gyr old or considerably more metal-rich, and others must be extremely dusty.  
The latter are part of a ``$J$-band dropout'' population which includes a number of
galaxies in one field which are probably at $z$$\sim$2.5, background to the quasar.
Finally, I discuss narrowband redshifted H$\alpha$ imaging in two fields 
which has yielded at least one detection.  
At $z$$\gtrsim$1.3 cluster galaxy star formation rates can be measured
from H$\alpha$ in the near-IR, rest-frame 1500\AA\ light in the optical,
and eventually rest-frame far-IR emission using SCUBA, to help determine
the star formation histories of cluster galaxies as function of redshift.}
\end{abstract}

\section{Introduction}
Studies of early-type cluster galaxies 
out to $z$$\sim$1 have shown that the vast majority 
are relatively old and red at all $z$$\leq$1, implying formation
redshifts for their dominant stellar populations of at least
$z$$>$2.5, and probably $z$$>$4 (\cite{bow98}).
Hierarchical clustering models predict that some star formation
(mostly merger-induced) in these galaxies continues down to 
at least $z$$\sim$1 (\cite{kau96}).
In contrast, monolithic collapse models predict a star formation epoch
of shorter duration due to expulsion of the ISM by supernovae (\cite{ay87}).
One way to test these somewhat competitive models is to study 
the properties of $z$$>$1 cluster galaxies, especially early-types;
e.g. their spectral energy distributions, star formation rates, etc.

Combined optical and near-IR imaging can easily identify $z$$>$1 clusters 
(although spectroscopic confirmation is difficult).
However, clusters of galaxies are rare objects and wide-field deep
near-IR imaging surveys are only now becoming feasible.  Another way to
find $z$$>$1 clusters efficiently is to target fields
where there is some reason to believe a cluster exists,
such as the presence of X-ray emission (\cite{ros98}),
quasar absorption line overdensities (\cite{fwd97}),
or a known high-redshift object (e.g. a radio galaxy, \cite{dic96b}).
This talk highlights results from a survey around radio-loud quasars (RLQs).
We targeted RLQs because at lower redshift they are known to often reside in 
clusters (\cite{ye93}), unlike radio-quiet quasars.

\section{Candidate Clusters around $z$=1--2 Radio-Loud Quasars}
Hall, Green \& Cohen (1998) and Hall \& Green (1998) describe and analyze
an $r$ and $K_s$ imaging survey of 31 RLQs at $z$=1--2.  The evidence for
clusters around these RLQs is as follows:

\noindent{$\bullet$~There is an excess
of $K$$\gtrsim$18.5 galaxies in these fields above the
average published literature counts (Fig.~1a).  The excess is significant even
when the relatively large field-to-field scatter in $K$ band counts (due to
the stronger clustering of red galaxies) is taken into account.}

\begin{figure}[t]
\plottwo{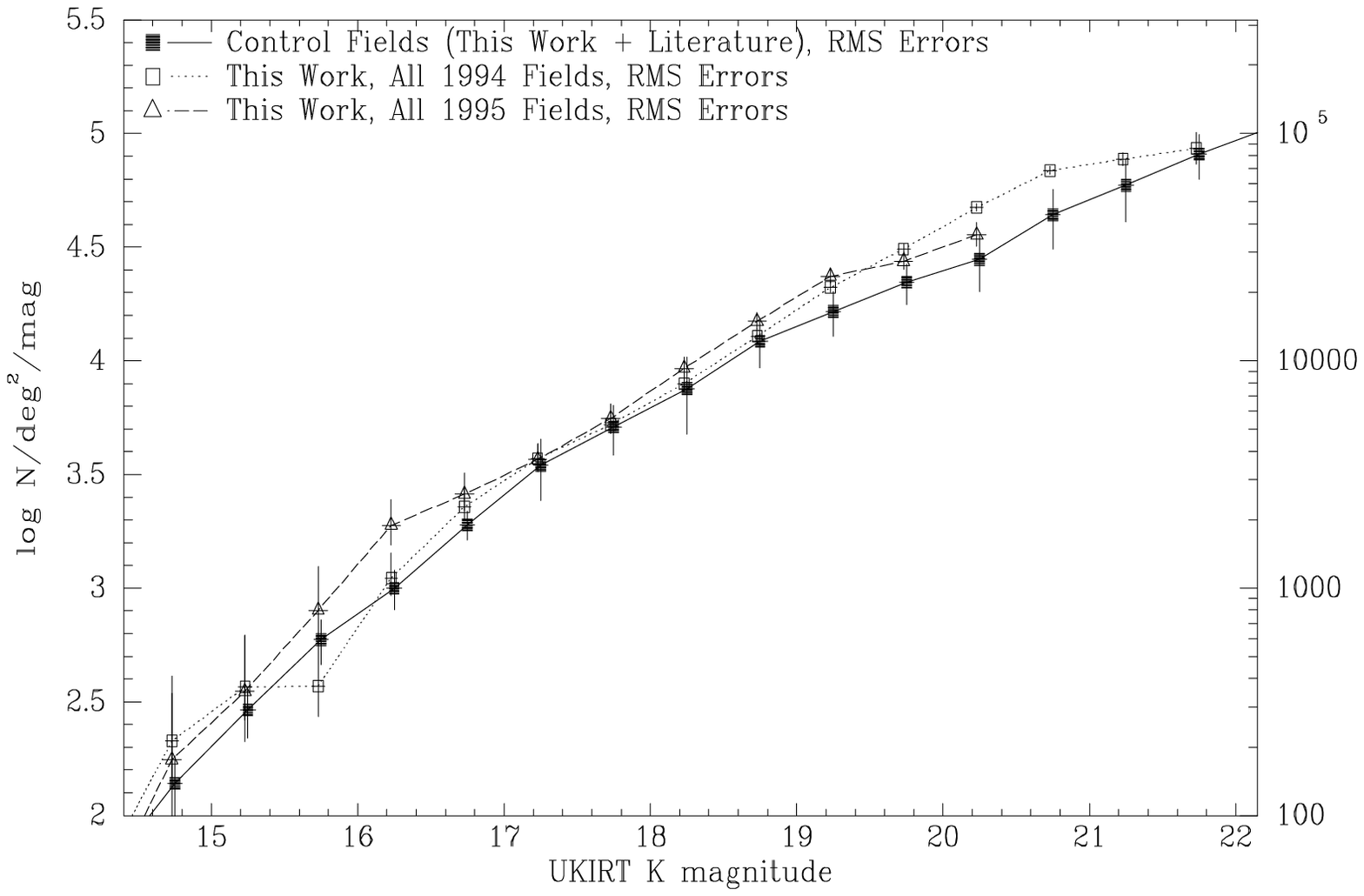}{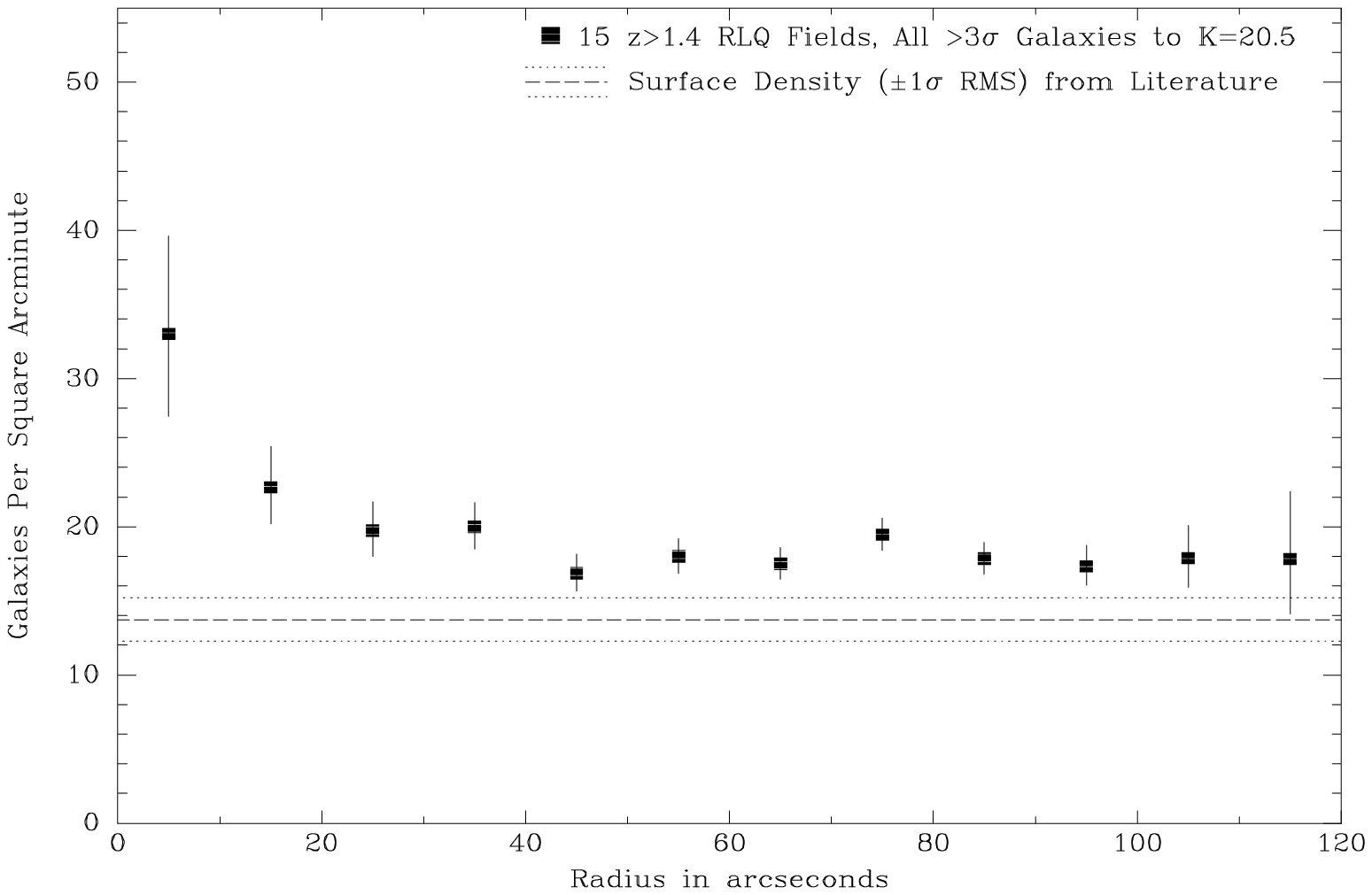}
\caption{
\singlespace
\small
{\bf a.}
The $K_{UKIRT}$ number-magnitude relations for quasar fields imaged during two
runs in 1994 and 1995 are plotted as dotted and dashed lines respectively.
The area-weighted average of our control fields and all published random-field
imaging surveys (corrected to $K_{UKIRT}$) is plotted as the filled points and
solid line, along with RMS errors on the N(m) values.
For clarity, 1995 fields are offset +0\fm06 and control fields $-$0\fm06.
Our magnitude scale has been conservatively adjusted faintwards to match the 
literature data at $K_{UKIRT}$=17--18.  Even after this adjustment, 
there is a significant excess in the quasar fields.
{\bf b.}
Radial distribution of galaxies relative to the 15 RLQs with the deepest imaging.
All galaxies detected at $\geq$3$\sigma$ down to $K$=20.5 are included.  Error bars
are calculated for the number of objects in each bin as per Gehrels (1986).  
At small radii the bin area is small, and at large radii only part of the full
annulus is imaged, so the uncertainties are large in both cases.
The dashed line is the surface density from the average published
literature counts; the dotted lines are the $\pm$1$\sigma$ RMS uncertainties.
There is a clear excess of galaxies at $<$40$''$ relative to $>$40$''$ and another
excess across the entire field compared to the literature data.
}\label{fig_knm_rp}
\end{figure}

\noindent{$\bullet$~There are two spatial components to the excess (Fig.~1b).
The first is a {\em near-field} excess within 40$''$ of the quasars relative to our
own data at $>$40$''$.  The second is a {\em far-field} excess relative to
random-field counts from the literature which extends to at least 100$''$.
The near-field excess is present in $\gtrsim$25\% of the fields, and the
far-field excess in $\gtrsim$50\%.
The amplitude of the near-field excess is insufficient to explain the excess
$K$ band counts, so we believe the far-field excess is real.

\noindent{$\bullet$~The color distribution of faint galaxies in these RLQ fields
is significantly redder than in random fields (Fig.~2).  
The excess galaxies essentially all have $r$$-$$K$$>$4,
redder than evolutionary synthesis models
of galaxy types Sa and later at any $z$=0--3 (\cite{pog97}).
The `typical' excess galaxy color is $r$$-$$K$$\sim$5.5,
redder than the Poggianti (1997) model ellipticals (Es) at any $z$=0--3.
This is because these model Es have exponentially declining star formation rates
(SFRs) with $\tau$$\geq$1~Gyr; models with faster declining SFRs match
the observed colors better (e.g. 1~Gyr burst model Es can have 
				 $r$$-$$K$$\gtrsim$5.5 at $z$$\gtrsim$0.9).
Thus, the $r$$-$$K$ color distribution of the excess galaxies is consistent with
a population of predominantly early-type galaxies at $z$$\gtrsim$0.9.}

\begin{figure}[t]
\epsscale{0.7}
\plotone{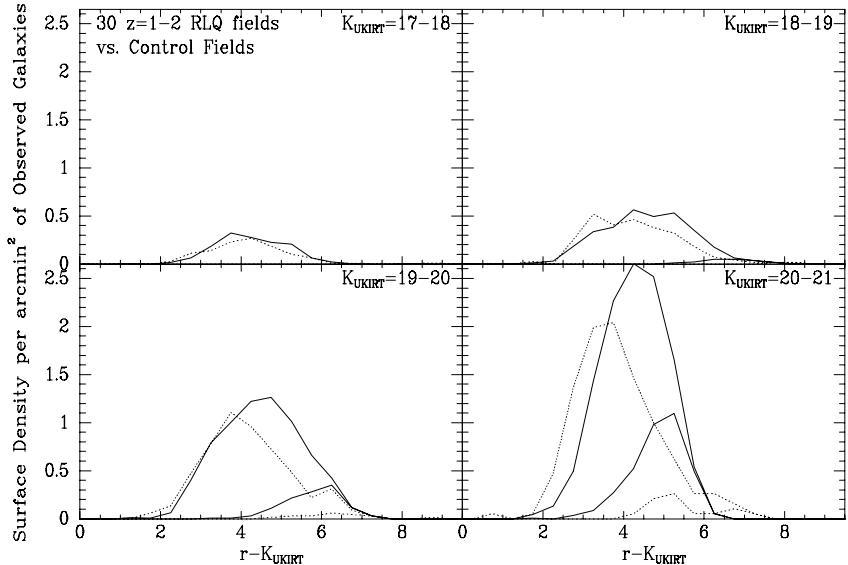}
\caption{
\singlespace
\small
Histograms of the surface densities of $K$-selected galaxies as a function of
$r$$-$$K$ color.
Solid lines are our $z$=1--2 quasar fields; dotted lines are control fields.  
Smaller histograms represent galaxies with lower or upper limits to their colors.
The $K_{UKIRT}$ magnitude range is given in each panel.  There is a statistically
significant excess of galaxies with $r$$-$$K$$>$4 at $K$$\gtrsim$19.
The deficit of blue galaxies at $K$=20--21 in the quasar fields is hard to
understand as real or an artifact (\cite{hg98}), but the red galaxy excess
at $K$=20--21 is still significant even if the histograms are arbitrarily
shifted to remove the blue galaxy deficit.
}\label{fig_krk}
\end{figure}

\noindent{$\bullet$~The 
excess is uncorrelated with the presence of
intervening Mg~{\sc ii} absorption 
at $z$$>$0.9.}

In summary,
the magnitude, spatial, and color distributions of the excess galaxy population
in these RLQ fields are all consistent with the excess being produced by 
overdensities of galaxies at the quasar redshifts.  
Roughly speaking, the amplitude of the near-field excess corresponds to Abell
richness class 0$\pm$1 clusters (where $-$1 denotes the richness of the field),
and the far-field excess to Abell richness 1.5$\pm$1.5.
This is consistent with RLQs often being located in large-scale galaxy overdensities
and occasionally in small-scale ``condensations'' within them.
However, the richness measurements should be 
interpreted cautiously, 
as one galaxy at $z$=0 may typically correspond to several galaxies at $z$$\sim$1.5
due to mergers between then and now (\cite{ste97}).

\section{Properties of Candidate Cluster Galaxies at $z$=1.5}

\begin{figure}[t]
\plottwo{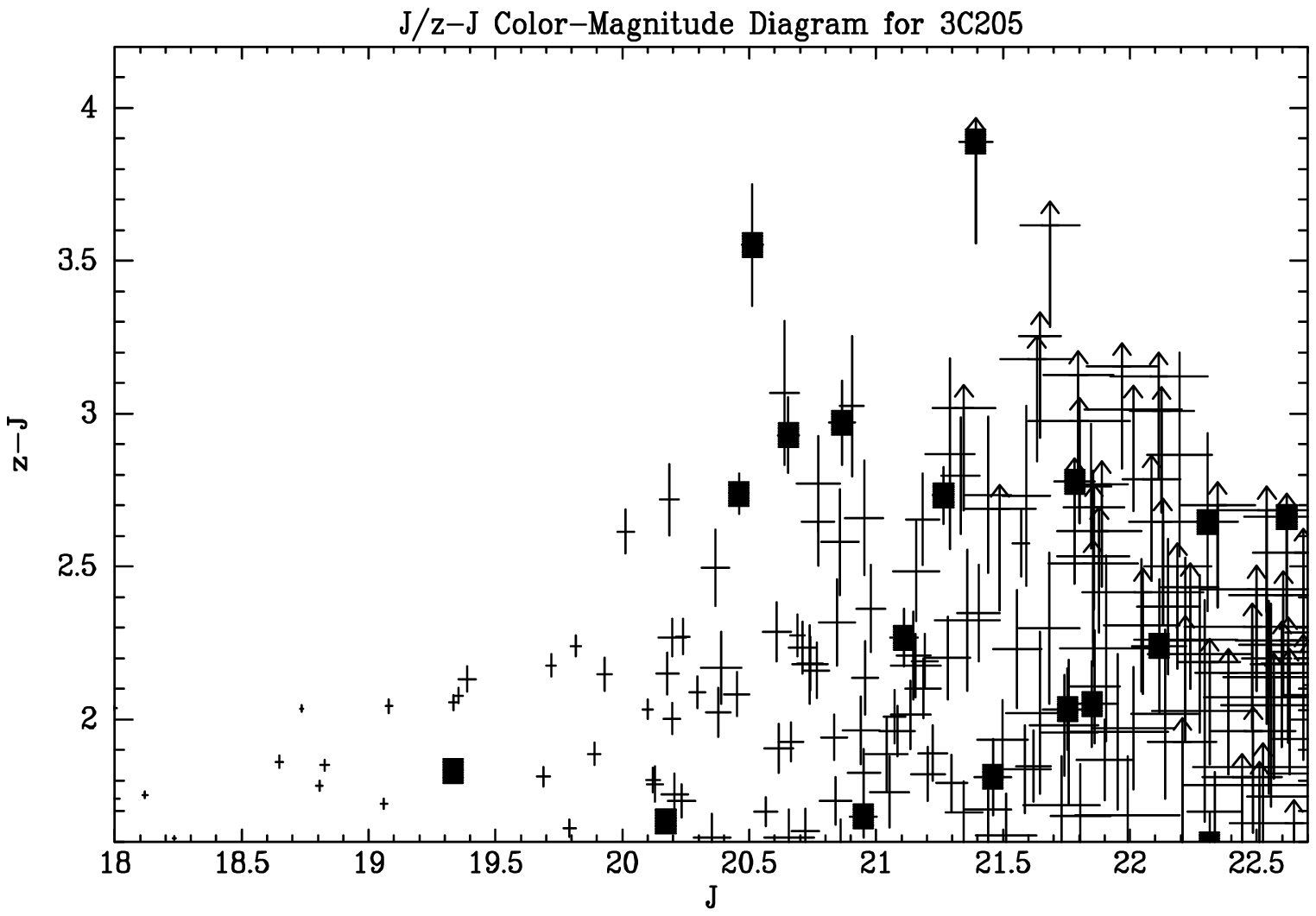}{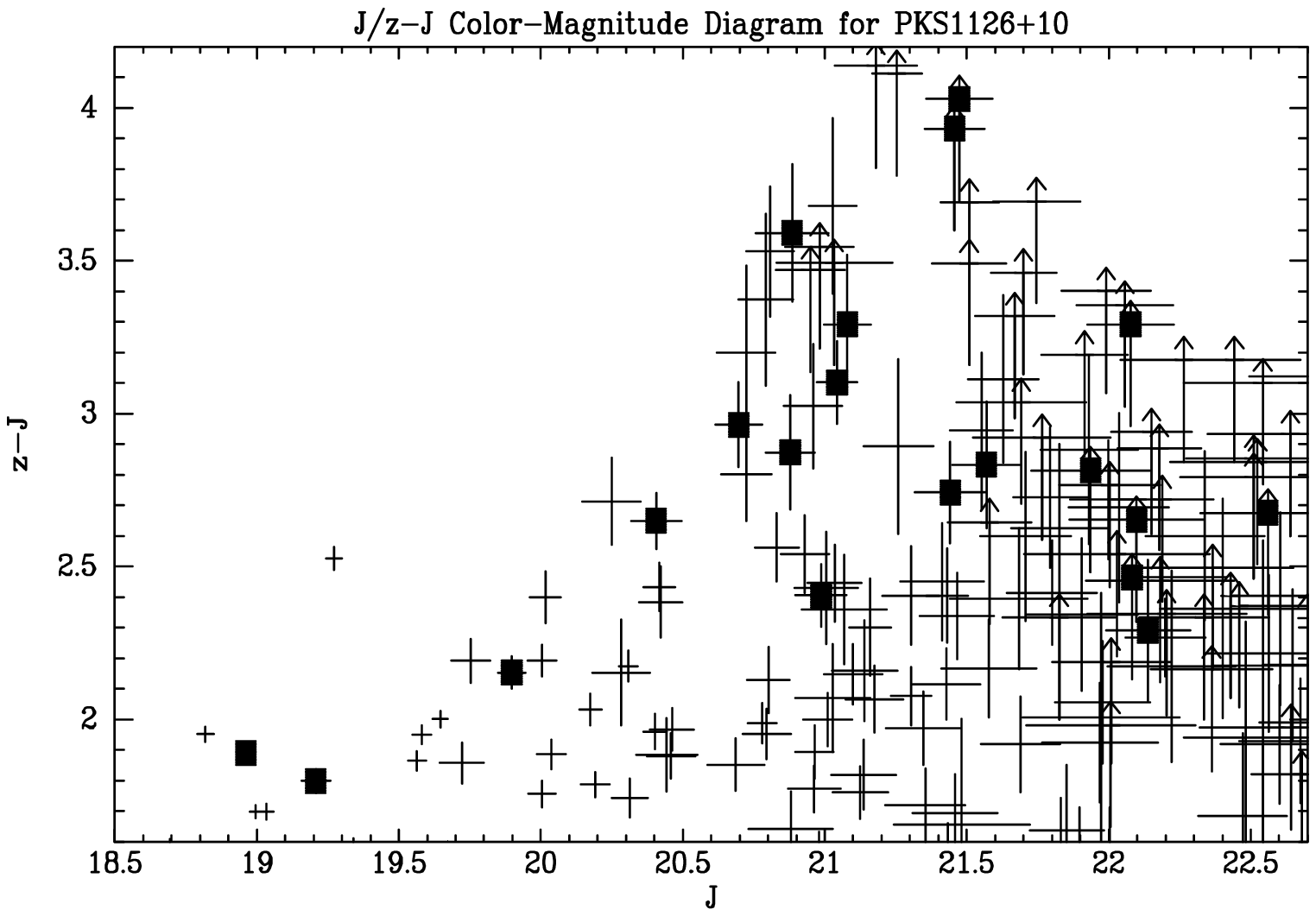}
\caption{
\singlespace
\small
J/z-J color-magnitude diagrams.  
{\bf a.} 3C205 field.
Filled boxes are objects within 20$''$ of 3C205.
The possible red sequence objects at z-J$\sim$2.8 have r-K$>$5
and 1.6$<$J-K$<$2.1, consistent with 2--3~Gyr galaxies at $z$=1.5;
the two z-J$\geq$3.5 objects have similar J-K but redder r-K,
consistent with 4--5~Gyr old galaxies at $z$=1.5.
{\bf b.} PKS1126+10 field.
The many ``$J$-dropouts'' in this field can be seen at z-J$\geq$3.
Filled boxes are objects with r-K$>$5 and 1.6$<$J-K$<$2.1.
As in the 3C205 field, there may be a red sequence at z-J$\sim$2.8,
along with a number of possibly older and/or dusty galaxies with z-J$>$3.5.
}\label{fig_jzj}
\end{figure}

\begin{figure}
\epsscale{1.00}
\plottwo{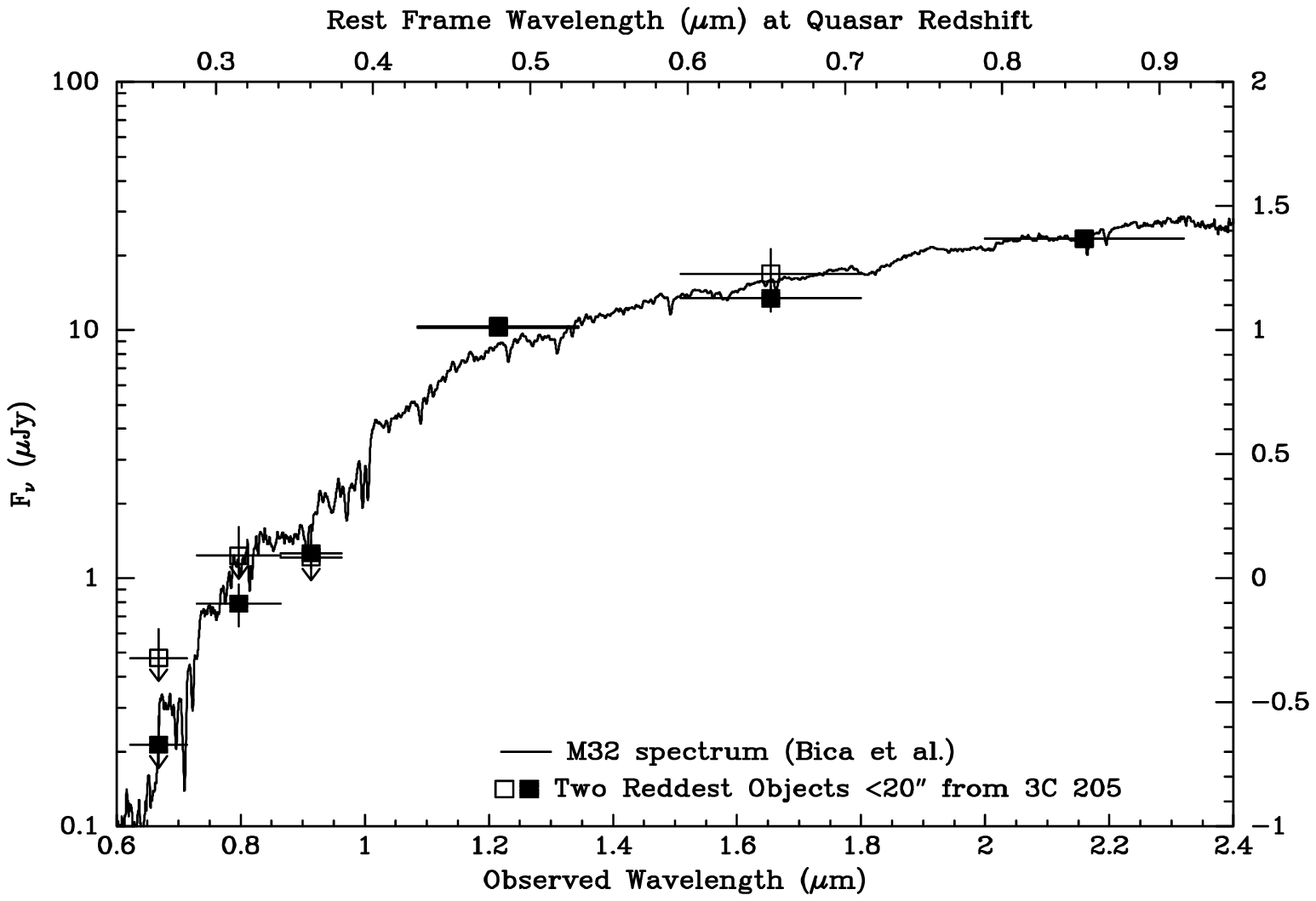}{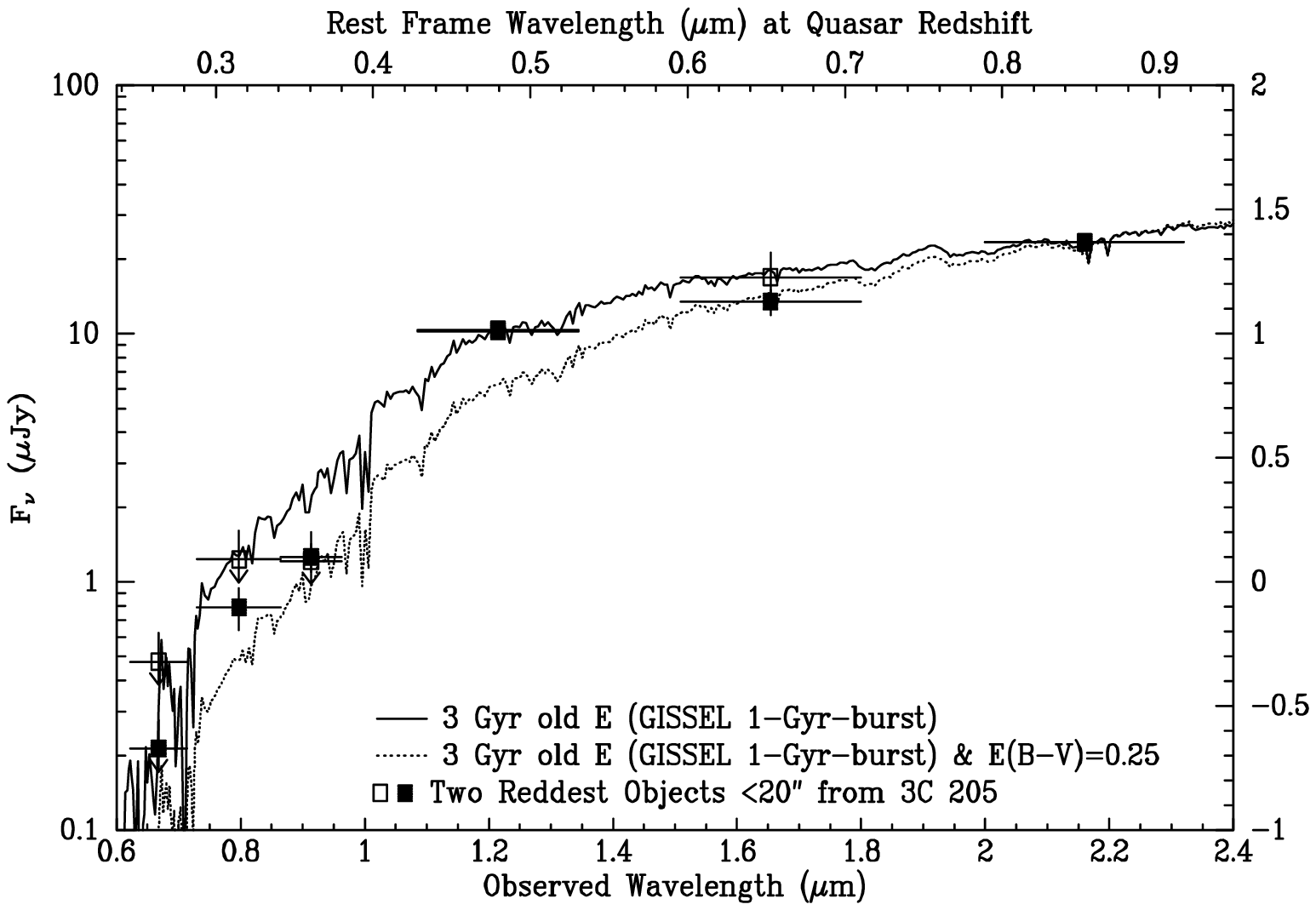}
\caption{
\singlespace
\small
The SEDs for the two reddest objects in $r$$-$$K_s$ within $\theta$=20\farcs5 of 
Q~0835+580 ($z$=1.534) are shown as the filled and open squares with Poisson error
bars on the fluxes.  Horizontal error bars indicate the widths of the filters used
to construct the SEDs.  
{\bf a.} The Bica {\em et al.} (1996) spectrum of M32 is shown as the solid line.
{\bf b.} Bruzual \& Charlot (1996) GISSEL 1-Gyr burst models viewed 3~Gyr
after the {\em start} of the burst are plotted here.  
The solid line is an unreddened spectrum and the dotted line is 
a dust-reddened spectrum, $E$($B$$-$$V$)=0.25.
The strong break between $z$ and $J$ rules out dust as the source of the red color.
}\label{fig_sed0835_m32dust}
\end{figure}

\begin{figure}
\epsscale{1.00}
\plottwo{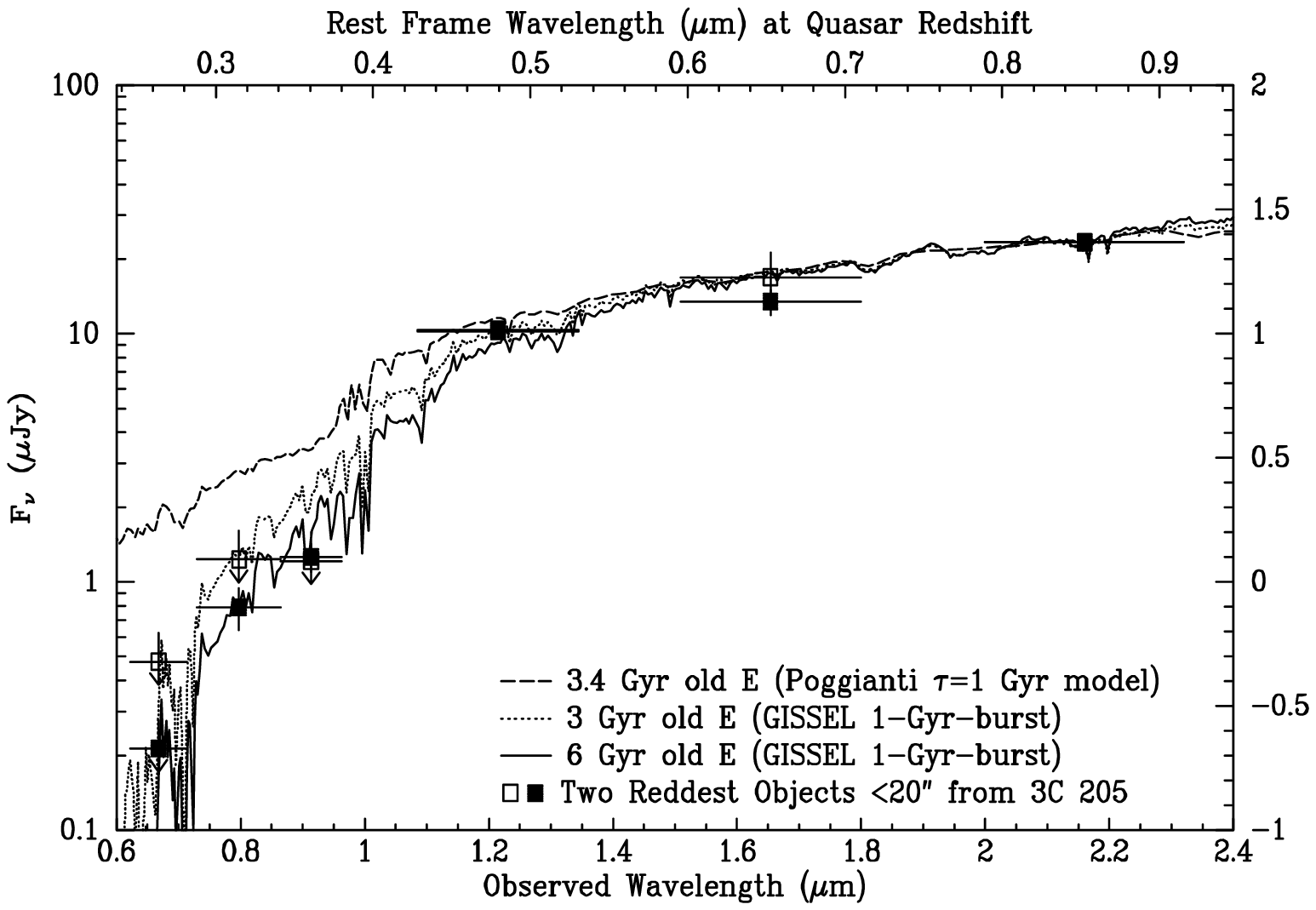}{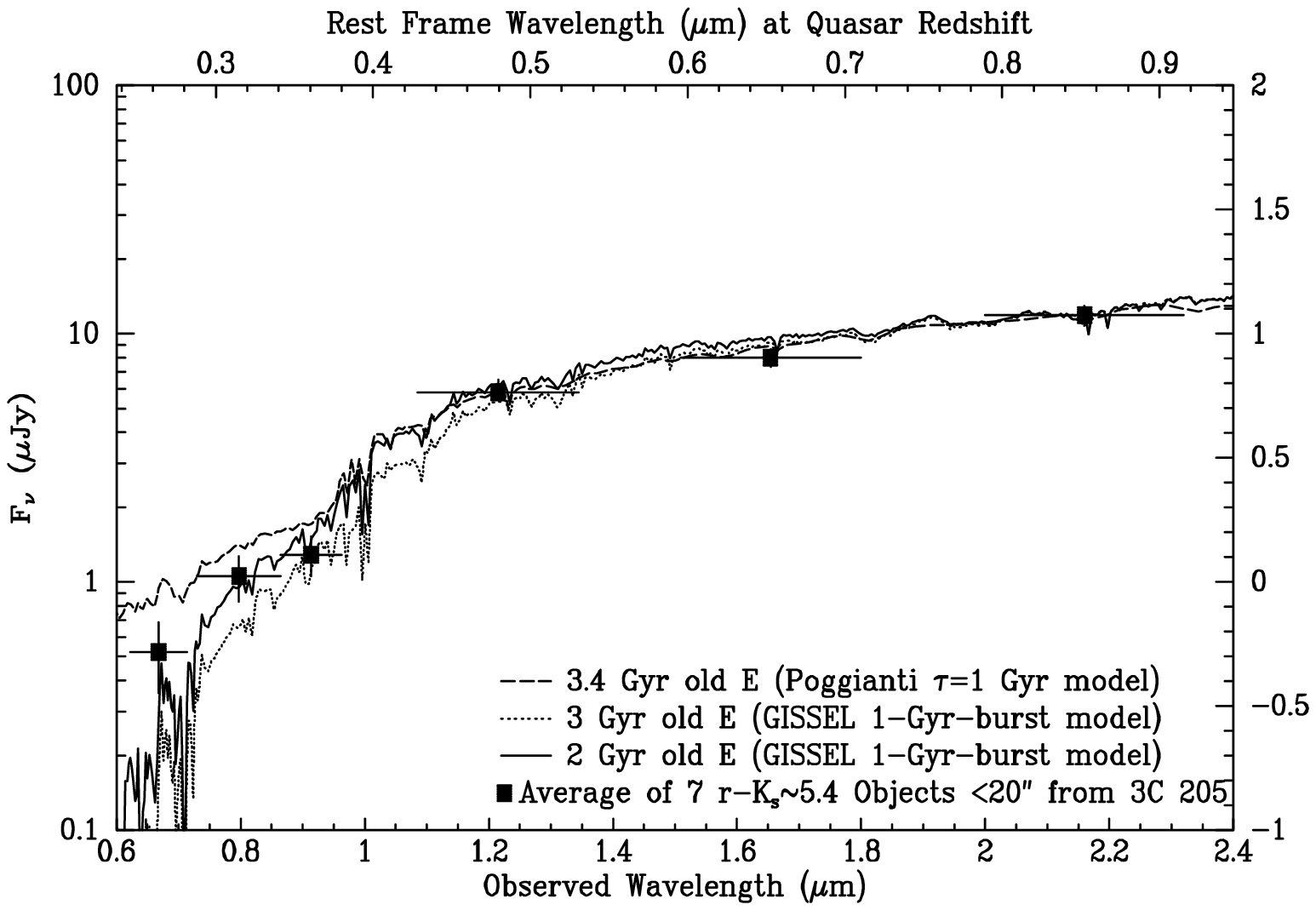}
\caption{
\singlespace
\small
{\bf a.} SEDs for the two reddest objects in $r$$-$$K_s$ within $\theta$=20\farcs5
of Q~0835+580 compared to 3- and 6-Gyr old GISSEL 1-Gyr-burst models and a 3.4~Gyr
old Poggianti (1997) model E with an exponentially declining SFR and $\tau$=1~Gyr.
The break between $z$ and $J$ rules out ages $\leq$3~Gyr.
{\bf b.} Average SED for the seven next reddest objects in $r$$-$$K_s$ within
$\theta$=20\farcs5 from Q~0835+580 compared to 2- and 3-Gyr old GISSEL 1-Gyr-burst 
models and the same Poggianti model E.
The break between $z$ and $J$ requires ages of 2$-$3~Gyr.
}\label{fig_sed0835_2v7}
\end{figure}

We do not yet have spectroscopy for these candidate cluster galaxies which would
enable detailed study of their properties, as well as confirm the reality of the
clusters.
However, we have imaging data in multiple bands for several fields,
notably Q~0835+580 (3C~205) and Q~1126+101 (both at $z$$\sim$1.5; color pictures 
available at http://www.astro.utoronto.ca/$\sim$hall/thesis.html).
If we assume that the bulk of the excess galaxies are at the quasar redshifts,
from these data we can place interesting constraints on the star formation
histories of cluster galaxies at $z$$\sim$1.5.

There is a distinct clump of galaxies within $\sim$20$''$ of 3C~205.
The $J$/$z$$-$$J$ color-magnitude diagram of these galaxies (Fig.~\ref{fig_jzj}a)
hints at a red sequence with $z$$-$$J$$\sim$2.8.  The uncertainties on the
colors are large due to the optical faintness of the galaxies, but there are 
two galaxies significantly redder than the others.  
In Fig.~\ref{fig_sed0835_m32dust}a,
the spectral energy distributions (SEDs) of these galaxies are compared to that
of M32, whose youngest stellar population is believed to be $\sim$4--5~Gyr old
(\cite{spi97}).  The match is excellent; these galaxies
are in fact redder (and thus possibly older) than the $z$=1.552 mJy radio source
LBDS 53W091, which may be old enough to significantly constrain the cosmological
model given the large lookback time to $z$=1.5 (\cite{spi97}).
Dust cannot explain the redness of these objects due to the strong break at 1$\mu$m
observed (Fig.~\ref{fig_sed0835_m32dust}b), consistent with the 4000\AA\ break
at the quasar $z$=1.5358.
(3C~205 does have two intervening Mg~{\sc ii} systems at $z$$=$1.437, but at least
the fainter of these two extremely red galaxies --- an $riz$-band dropout ---
is likely to be at the quasar $z$ instead of the Mg~{\sc ii} $z$, 
since it is positionally coincident with the 
compact radio hot spot which terminates one of the quasar's radio jets;
cf. \cite{lb98}.)
In Fig.~\ref{fig_sed0835_2v7}a the same SEDs are compared to Bruzual \& Charlot
(1996) GISSEL 1-Gyr-long burst model ellipticals and to a 3.4~Gyr old Poggianti 
(1997) model elliptical (all with solar metallicity).
The $\tau$=1~Gyr exponentially declining SFR of the Poggianti model
is bluer than the data, and remains bluer even when 5.9~Gyr old (not shown),
whereas a 6~Gyr old GISSEL model is a good fit.
Thus models with $\tau$$\gtrsim$1~Gyr exponentially declining SFRs
are strongly disfavored for at least the reddest candidate cluster ellipticals.
Fig.~\ref{fig_sed0835_2v7}b plots the average SED of the seven next reddest
galaxies near 3C~205 (the putative red sequence in Fig.~\ref{fig_jzj}a).
2--3~Gyr GISSEL models can match the data
(with a bit of later star formation to explain the $r$ fluxes)
as can a 4.3~Gyr old Poggianti E (not shown).
The difference in color between these candidate early-type cluster galaxies implies
a difference in age of $\gtrsim$2--3~Gyr or in metallicity of $\gtrsim$0.2~dex.
If real, this large scatter must be erased at later times through merging
to explain the small scatter seen in the red sequence at lower $z$.

\begin{figure}[t]
\plottwo{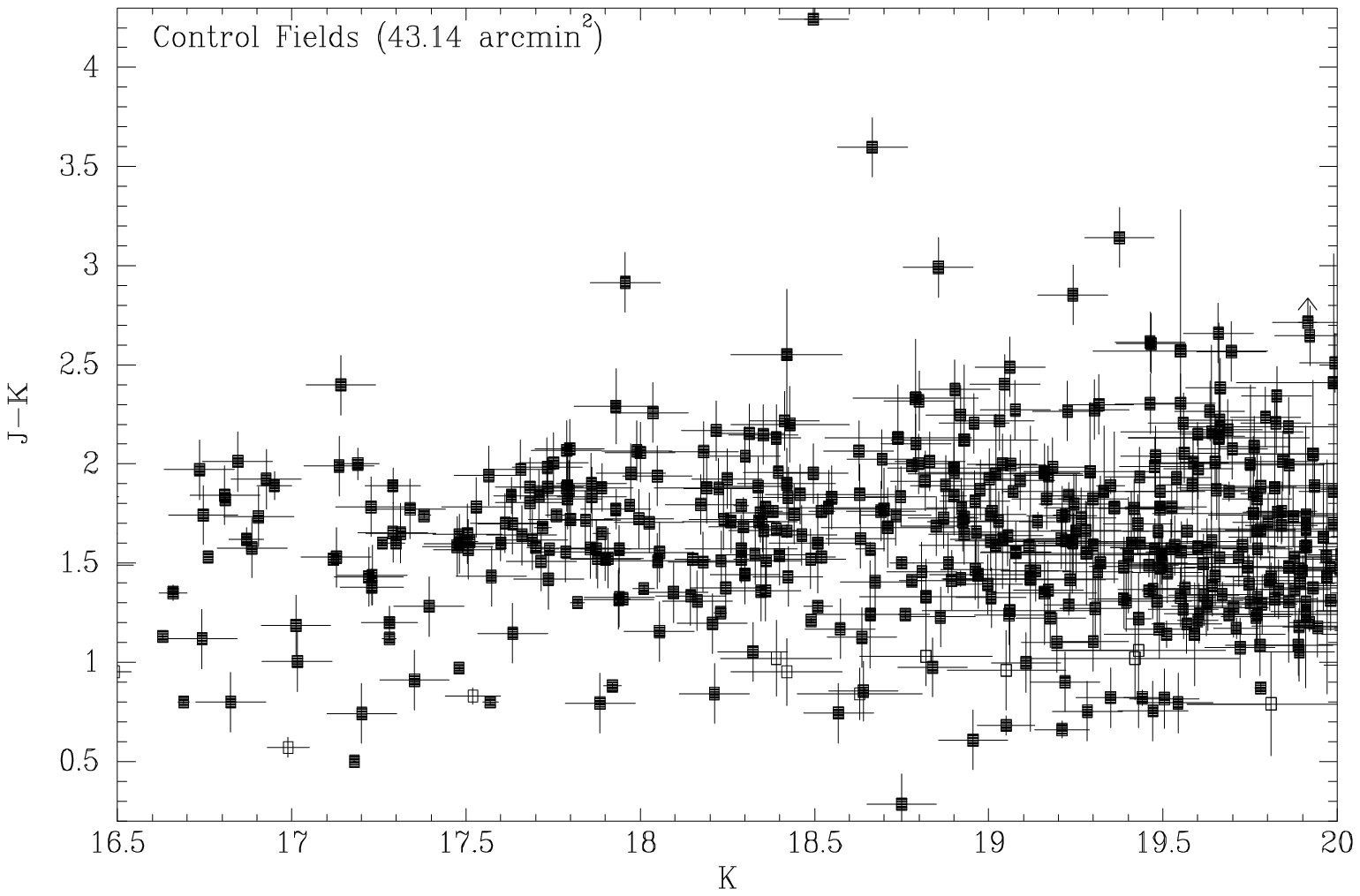}{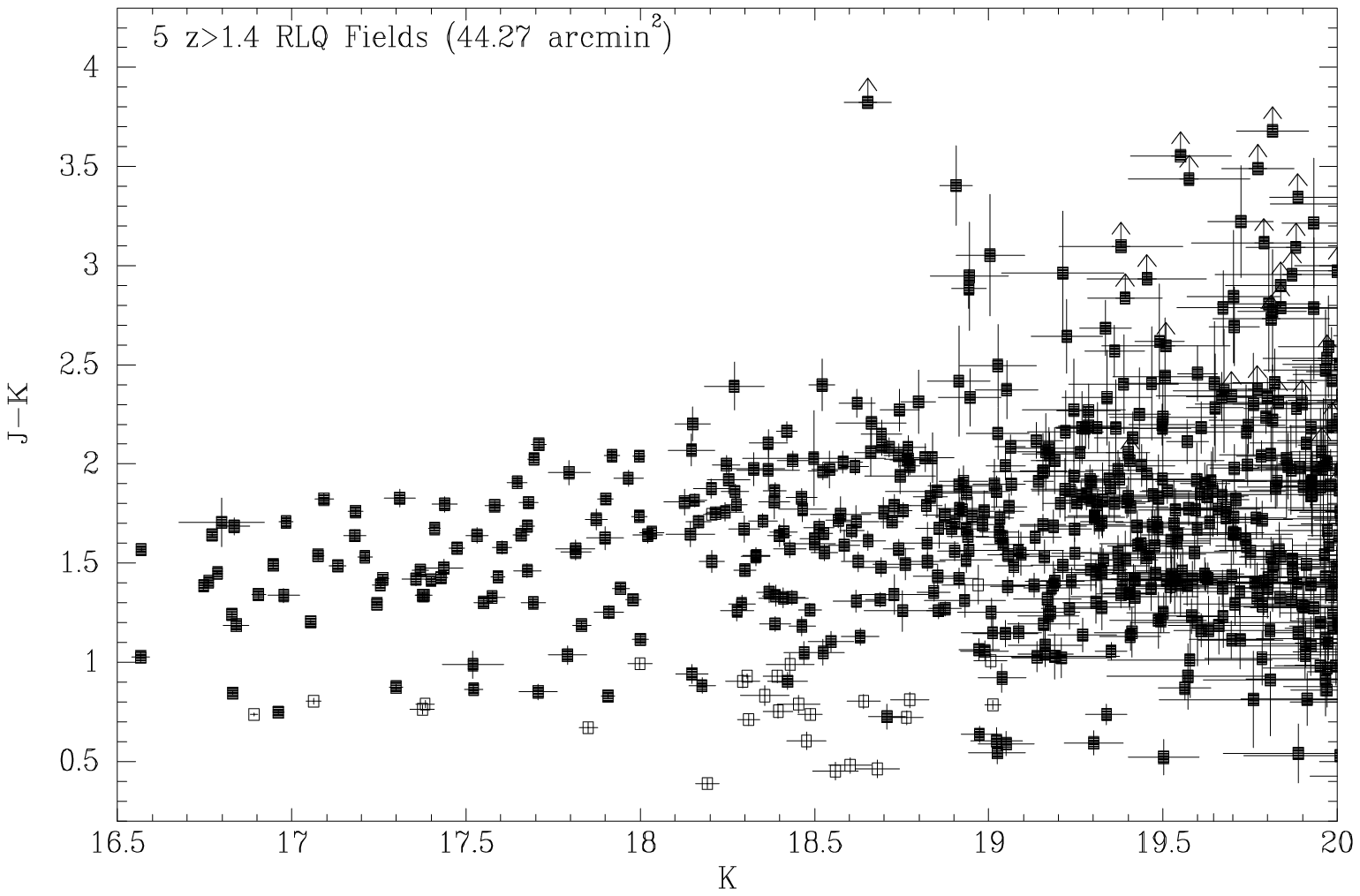}
\caption{
\singlespace
\small
$K$/$J$$-$$K$ color-magnitude diagrams.
Open symbols are stars and filled symbols are galaxies.
{\bf a.} Galaxies in the control field datasets of McLeod {\it et al.} (1995),
Elston {\it et al.} (1998) and Dickinson {\it et al.} (1998).  
{\bf b.} Galaxies in our five $z$=1.4--2 RLQ fields with $J$ data.  
There is a clear excess at $K$$>$19, $J$$-$$K$$>$2.5.
}\label{fig_kjk}
\end{figure}

\begin{figure}[t]
\epsscale{1.05}
\plotone{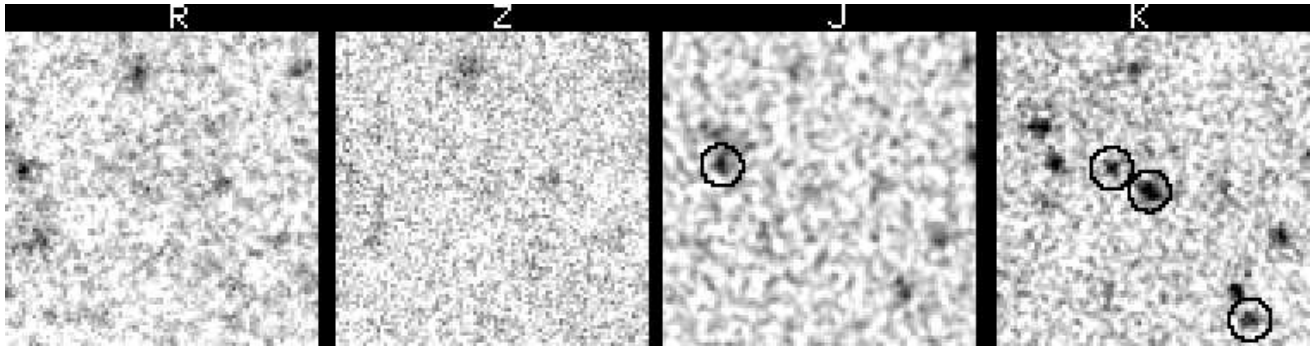}
\caption{
\singlespace
\small
Several ``$J$-dropouts'' ($J$$-$$K_s$$>$2.5) in the
Q1126+101 field, circled in $K_s$.
The central object appears extended in $K_s$ and is detected at $r$$\sim$25.
The object circled in $J$ is a $z$-dropout, but is strongly detected in $J$.
Its SED is consistent with a 4--5~Gyr old galaxy at the quasar redshift.
}\label{fig_zdrop}
\end{figure}

\begin{figure}[t]
\epsscale{1.00}
\plottwo{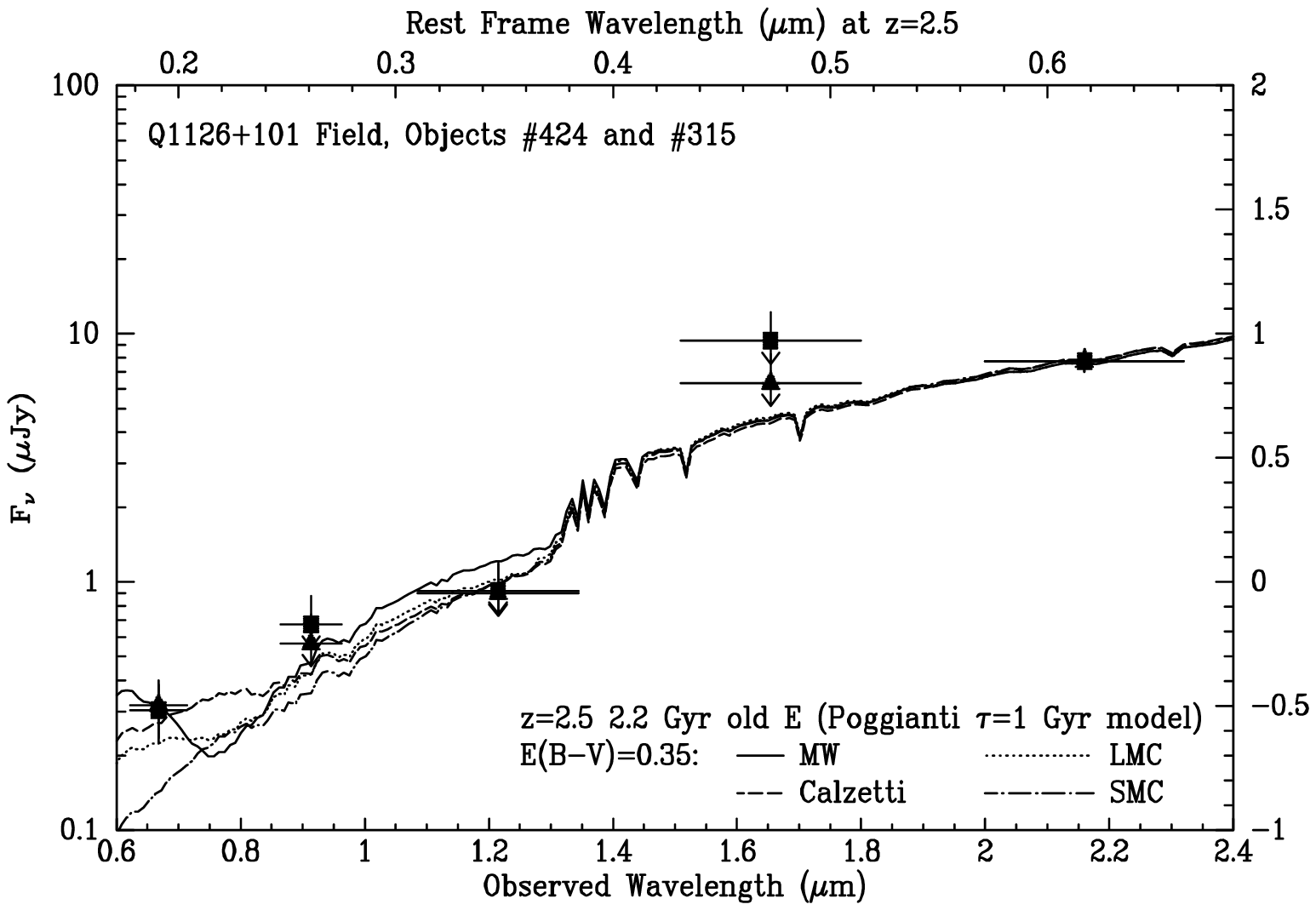}{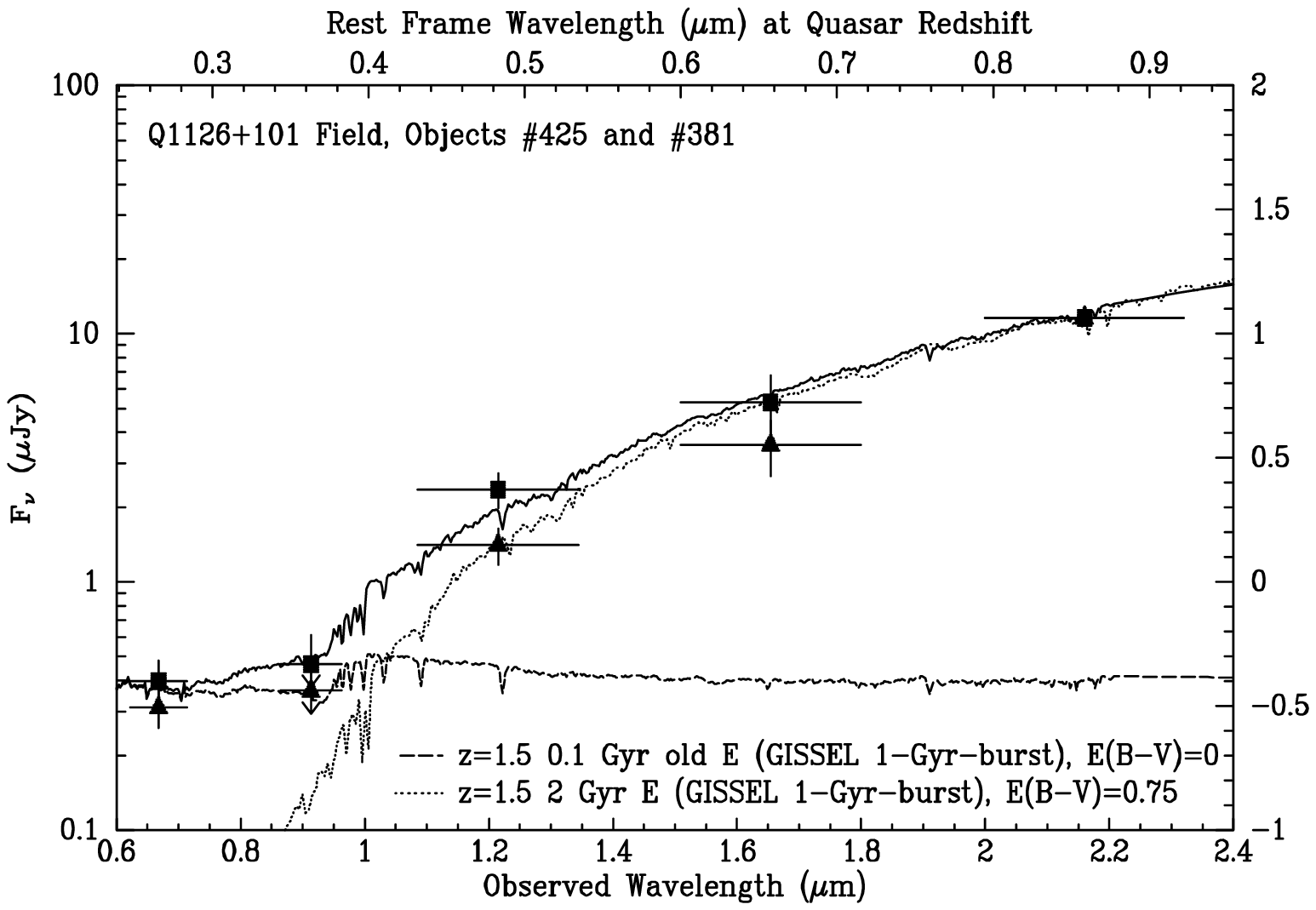}
\caption{
\singlespace
\small
{\bf a.} 
Points are the two objects with $r$$-$$K_s$$>$5 which are reddest in
$J$$-$$K_s$ in the Q~1126+101 field, normalized to the same flux in $K_s$.
Lines are the 2.2~Gyr old E model of Poggianti (1997) reddened by different
extinction laws with $E(B$$-$$V)$=0.35 and normalized to the data in $K_s$.
The extinction laws are Milky Way (MW; solid line), Calzetti (1997; dashed),
LMC (dotted) \& SMC (dash-dot).  
The relative extinctions are significantly different only
at rest-frame $\lambda$$<$3800~\AA.
{\bf b.} Data points are the two objects with $J$$-$$K_s$$>$2.5 in the Q~1126+101
field which have believable detections in $H$, normalized to the same flux in $K_s$.
The dotted line is a 2~Gyr old GISSEL
model E reddened by $E(B$$-$$V)$=0.75 using the extinction law from Calzetti (1997)
and normalized to the data in $K_s$.  The dashed line is a 0.1~Gyr old unreddened
GISSEL model E normalized to the $r$ data.  
The solid line is the sum of the two, and it fits the data reasonably well.
}\label{fig_sed1126}
\end{figure}

There is no dramatic clump of galaxies around Q~1126+101, but galaxies with 
$r$$-$$K$ and $J$$-$$K$ colors similar to those in the 3C~205 clump show a possible
red sequence in this field as well, though it is less obvious (Fig.~\ref{fig_jzj}b).
Again, there are several galaxies much redder than the putative red sequence, and
some of these are $z$-band dropouts consistent with 4--5~Gyr old galaxies at the
quasar redshift ($z$=1.5173).  However, there are many more objects with very red 
$z$$-$$J$ colors in this field, and many of these objects have very red $J$$-$$K$
colors as well (Fig.~\ref{fig_kjk}).  We denote objects with $J$$-$$K$$\geq$2.5 
as $J$-band dropouts (Fig.~\ref{fig_zdrop}); typically they are $i$- and $z$-band
dropouts as well, although some are detected in $r$.
Some of these objects show a strong break only between $J$ and $K_s$ 
(Fig.~\ref{fig_sed1126}a);
they are best explained as moderately dusty galaxies background to the quasar.
Others show strong breaks between $J$ and $K_s$ {\em and} $z$ and $J$ 
(Fig.~\ref{fig_sed1126}b);
they are best explained as extremely dusty galaxies at the quasar redshift.
Some of these latter objects are redder in $J$$-$$K$ than the prototypical dusty
starburst HR10 at $z$=1.44 (\cite{gd96}), whose colors must
be due to dust rather than age for any reasonable cosmology.
The surface density of $J$-band dropouts with $r$$<$25.5 is
2.80/\sq\arcmin\ in the Q~1126+101 field, 1.29/\sq\arcmin\ 
in our other fields, and 0.46/\sq\arcmin\ in several field surveys.
These surface densities and SED fits suggest that about half these objects
are at the quasar redshifts and half are background to the quasars.

The existence of $J$-dropouts which are probably at $z$$\gtrsim$2.5 has important
implications for any census of galaxy populations at high redshift.
Galaxies at $z$=2.5--6 are now routinely
identified via their redshifted far-UV Lyman breaks, but such galaxies
must have considerable rest-frame far-UV flux to be found this way.
As pointed out by Francis {\it et al.} (1997), since both dust and age can reduce a 
galaxy's UV flux, 
there may be a population
of $z$$>$2.5 galaxies which are undetectable via the Lyman-break technique,
particularly since even rest-frame UV-selected galaxies can be dusty (\cite{sy98}).
Francis {\it et al.} (1997) have in fact spectroscopically confirmed three 
galaxies at $z$=2.38, selected via absorption lines in nearby background quasars,
whose red colors require ages $\geq$2~Gyr or dust reddenings $E(B$$-$$V)$$>$0.5.
These and other red galaxies nearby have similar bolometric luminosities to 
UV-selected objects, and along with the objects in our fields they may constitute a
population of previously overlooked dusty and/or old galaxies at $z$$\gtrsim$2.5.

The dispersion in SED fits to candidate cluster galaxies, probably early type
galaxies for the most part, implies a dispersion in their star formation histories.
A similar result has been found for another 
candidate quasar host cluster at $z$=1.1 (\cite{yam98}).
Most candidate cluster galaxies have little ongoing star formation, but a few
objects must be significantly older or more metal-rich than the others.
Some objects must be very dusty; if at the quasar $z$ they are best fit by
an actively star-forming population plus a $\sim$2~Gyr old component.

\section{H$\alpha$ Emission in Two Candidate Clusters at $z$=1.5}

\begin{figure}[t]
\epsscale{0.80}
\plotone{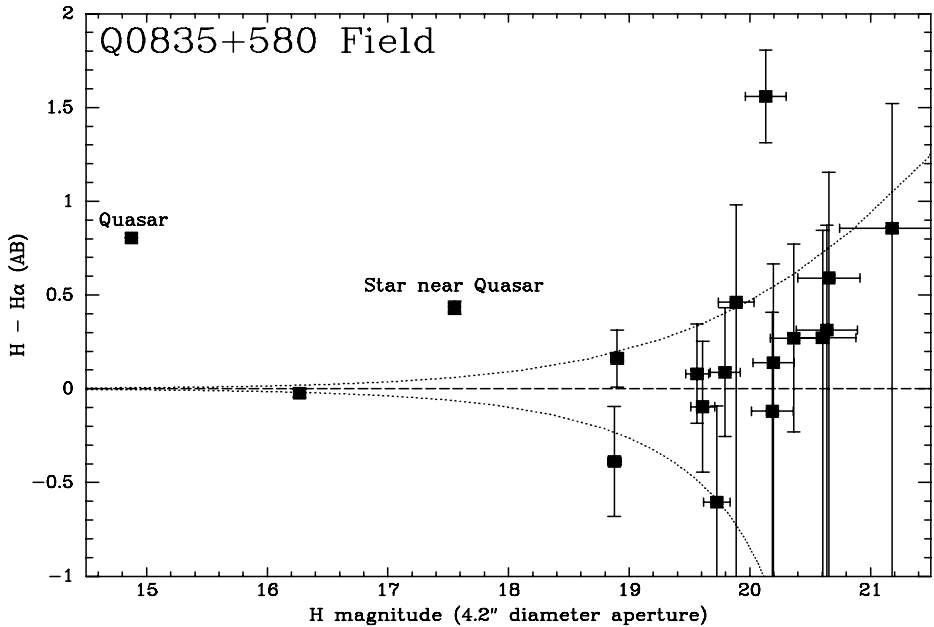}
\caption{
\singlespace
\small
This figure shows a candidate H$\alpha$ emitter at the redshift of Q0835+580.
The abscissa is $H$ magnitude and the ordinate is $H$$-$$H\alpha$ color 
in AB magnitudes, where $H\alpha$ is the narrowband magnitude at 1.6642$\mu$m,
corresponding to the wavelength of H$\alpha$ at the quasar redshift.
Filled squares and $\pm$1$\sigma$ error bars are shown for all objects in the
1\sq\arcmin\ field with narrowband IRTF data; error bars
extending off the lower edge graph denote 2$\sigma$ upper limits in $H\alpha$.
The dotted lines show the $\pm$3$\sigma$ detection significance criteria
as a function of $H$ magnitude.
The quasar and a nearby star contaminated by the quasar's PSF show excess
narrowband emission.  Only one galaxy shows an excess at $>$3$\sigma$,
with f$_{H\alpha}$=1.39~10$^{-16}$ ergs~s$^{-1}$~cm$^{-2}$.  
}\label{fig_ha}
\end{figure}

The NASA Infra-Red Telescope Facility's NSFCAM instrument has a circularly variable
filter which can be tuned to redshifted H$\alpha$ at $z$=1.3--2.83.
In October 1997 we imaged two $z$$\sim$1.5 candidate cluster fields to search for
H$\alpha$ emission.  The field of view is only 1\sq\arcmin\ in each cluster, but
in the 3C~205 field we detected one galaxy at $\sim$6$\sigma$ (Fig.~\ref{fig_ha}).
The galaxy's faint magnitude ($H$=20.1), blue $r$$-$$z$ and reddish $z$$-$$J$
colors are consistent with it being at the quasar redshift, 
in which case its inferred SFR$_{H\alpha}$ is 14.7$\pm$3.7 $M_{\odot}$/yr
($H_0$=75 km~s$^{-1}$~Mpc$^{-1}$, $\Omega_M$=0.2, $\Omega_\Lambda$=0).
(Note that most strong emission lines are blueward of H$\alpha$, so if the excess
narrowband emission is not caused by H$\alpha$ the object is likely at $z$$>$1.5.)
We have also obtained $U$ band imaging of this field to sample rest-frame 1500\AA\ 
light from galaxies at the quasar redshift.  The candidate H$\alpha$ emitter has 
$U$=23.0, corresponding to SFR$_{FUV}$=5.6 $M_{\odot}$/yr uncorrected for dust.
Thus there is relatively good agreement between SFR$_{H\alpha}$ and SFR$_{FUV}$.
H$\alpha$ and $U$-band observations of more cluster fields will determine current
SFRs and limits for more cluster galaxies at $z$$\gtrsim$1.3.
We have also proposed to map several fields with SCUBA to determine SFR$_{FIR}$ and
to see if the recently discovered field population of
sub-mm-luminous galaxies (\cite{sma98}) also exists in $z$$\sim$1.5 clusters.

\section{Summary}


Optical/near-IR imaging of radio-loud quasars at $z$=1--2 has revealed evidence for
overdensities of red galaxies at the quasar redshifts.  
Qualitative SED fitting to broadband imaging data in multiple bandpasses yields
interesting possible constraints on the star formation histories and current star
formation rates of these candidate $z$$>$1 early-type cluster galaxies.
(These are being put on a more quantitative footing via calculation of photometric
redshifts and formal constraints on age, reddening, etc.,
in collaboration with M. Sawicki; cf. Sawicki \& Yee 1998.)
The dispersion in SED fits suggests that some of these galaxies are significantly
older or more metal-rich than the majority, and that some are very dusty and 
probably have considerable ongoing star formation given their $\sim$L$^*$ 
luminosities.  This dispersion 
is generally consistent with the late stages of a hierarchical merger-induced star
formation epoch in early-type galaxies in these clusters; however, extended star
formation histories are not favored for the reddest such galaxies at least.
There are also a number of very red galaxies which are best fit
as background galaxies at $z$$\gtrsim$2.5; these may be part of
a population of previously overlooked dusty and/or old high-redshift galaxies.
Lastly, we have also detected redshifted H$\alpha$ emission from at least one
candidate cluster galaxy.  

Spectroscopic confirmation of these objects' redshifts is a job for 8-m class
telescopes, but multislit IR spectroscopy on 4-m telescopes is feasible for the
brightest red objects, and we have been granted time for this on CFHT with OSIS-IR.
Verifying that there is a large scatter in age and/or metallicity among early-type
galaxies in clusters at $z$$>$1 will require redshifts and spectral types for the
candidate early-type galaxies, as well as morphologies from proposed deep WFPC2
imaging or near-IR adaptive optics observations.
Multi-wavelength observations can be used to constrain SFRs in cluster galaxies
at $z$$>$1.3 three different ways:  via H$\alpha$ in the IR, 
via rest-frame far-UV in the optical, and via the rest-frame far-IR with SCUBA.
These various observations should eventually enable us to determine the star
formation histories of all types of cluster galaxies at $z$$>$1 and 
the relative importance of hierarchical merging and monolithic collapses.


\acknowledgements
I thank the organizers of the
13th Kingston Meeting on Theoretical Astrophysics:
    ``Galaxy Formation and Cosmic Star Formation History'' for financial support.

{}


\small
\begin{thebibliography}{}
\bibitem[Arimoto \& Yoshii 1987]{ay87}\reference{}    
Arimoto, N., \& Yoshii, Y.  1987, \aap, 173, 23
\bibitem[Bica {\it et al.} 1996]{bic96} \reference{}  
Bica, E., Alloin, D., Bonatto, C., Pastoriza, M.~G., Jablonka, P., Schmidt, A.,
and Schmitt, H.~R.  1996, in ``A Data Base for Galaxy Evolution Modeling,''
eds. C. Leitherer {\it et al.}, \pasp, 108, 996
\bibitem[Bower {\it et al.} 1998]{bow98}\reference{}    
Bower, R. G., Terlevich, A., Kodama, T., and Caldwell, N. 1998, to appear in ``Star
Formation in Early-Type Galaxies," eds. P. Carral and J. Cepa (astro-ph/9808325)
\bibitem[Bruzual \& Charlot 1996]{bc96} \reference{}    
Bruzual A., G., and Charlot, S.  1996, in ``A Data Base for Galaxy Evolution
Modeling,'' eds. C. Leitherer {\it et al.}, \pasp, 108, 996
\bibitem[Calzetti 1997]{cal97} \reference{}
Calzetti, D.  1997, to appear in ``The Ultraviolet Universe at Low and High
Redshift," eds. W.~H. Waller, M.~N. Fanelli, and A.~C. Danks (AIP: New York)
(astro-ph/9706121)
\bibitem[Dickinson 1996]{dic96b}\reference{}    
Dickinson, M.  1996, in ``The Early Universe with the VLT," ed. J. Bergeron
(Springer-Verlag: Berlin), 274
\bibitem[Dickinson {\it et~al.} 1998]{dic98}\reference{}        
Dickinson {\it et~al.}  1998, in preparation
\bibitem[Elston, Eisenhardt \& Stanford 1998]{ees97} \reference{}
Elston, R., Eisenhardt, P., and Stanford, S.~A.  1998, in preparation
\bibitem[Francis, Woodgate \& Danks 1997]{fwd97} \reference{}
Francis, P.~J., Woodgate, B.~E., and Danks, A.~C.  1997, \apj, 482, L25
\bibitem[Gehrels 1986]{geh86}\reference{}
Gehrels, N.  1986, \apj, 303, 336
\bibitem[Graham \& Dey 1996]{gd96}\reference{}
Graham, J.~P., and Dey, A.  1996, \apj, 471, 720
\bibitem[Hall, Green \& Cohen 1998]{hgc98} \reference{}
Hall, P.~B., Cohen, M., and Green, R.~F.  1998, \apjs, 119, 1 (astro-ph/9806145)
\bibitem[Hall \& Green 1998]{hg98} \reference{}
Hall, P.~B., and Green, R.~F.  1998, \apj, 507, 558 (astro-ph/9806151)
\bibitem[Kauffmann 1996]{kau96} \reference{}
Kauffmann, G.  1996, \mnras, 281, 487 (astro-ph/9502096)
\bibitem[Lonsdale \& Barthel 1998]{lb98} \reference{}           
Lonsdale, C.~J., and Barthel, P.~D.  1998, \aj, 115, 895
\bibitem[McLeod {\it et~al.} 1995]{mcl95}\reference{}
McLeod, B.~A., Bernstein, G.~M., Rieke, M.~J., Tollestrup, E.~V.,
and Fazio, G.~G.  1995, \apjs, 96, 117
\bibitem[Poggianti 1997]{pog97} \reference{}
Poggianti, B.~M.  1997, \aaps, 122, 399
\bibitem[Rosati 1998]{ros98} \reference{}
Rosati, P., to appear in ``Wide Field Surveys in Cosmology'', eds. S. Colombi
and Y. Mellier (Gif-sur-Yvette: Editions Frontieres) (astro-ph/9810054)
\bibitem[Sawicki \& Yee 1998]{sy98} \reference{}        
Sawicki, M., and Yee, H.~K.~C.  1998, \aj, 115, 1329 (astro-ph/9712216)
\bibitem[Smail {\it et~al.} 1998]{sma98} \reference{}
Smail, I., Ivison, R., Blain, A, \& Kneib, J.-P.  1998, to appear in the proceedings
of ``After the dark ages: when galaxies were young (the universe at 2$<$$z$$<$5)''
(astro-ph/9810281)
\bibitem[Spinrad {\it et~al.} 1997]{spi97} \reference{} 
Spinrad, H., Dey, A., Stern, D., Dunlop, J., Peacock, J., Jimenez, R., \&
Windhorst, R. 1997, \apj, 484, 581
\bibitem[Steinmetz 1997]{ste97} \reference{}
Steinmetz, M.  1997, to appear in ``Structure and Evolution of the IGM From QSO
Absorption Line Systems'', eds. P. Petitjean and S. Charlot (Nouvelles
Frontieres: Paris) (astro-ph/9709260)
\bibitem[Yamada {\it et al.} 1998]{yam98} \reference{}
Yamada, T., {\it et al.}  1998, in {\em The Young Universe}, eds. S. D'Odorico
{\it et al.} 
(ASP: San Francisco), 526
\bibitem[Yee \& Ellingson\ 1993]{ye93}\reference{}
Yee, H.~K.~C., and Ellingson, E., 1993, \apj, 411, 43 

\end{thebibliography}
\end{document}